\documentclass[aip,apl,12pt,reprint,nofootinbib]{revtex4-1}

\usepackage{subfig}
\usepackage{graphicx}% Include figure files
\usepackage{dcolumn}% Align table columns on decimal point
\usepackage{bm}% bold math
\usepackage{ulem}
\usepackage{amsmath}
\usepackage{color}

\newcommand{\vect}[1]{\bm{#1}}

\begin{document}

\title{System for control of polarization state of light and generation of light with continuously rotating linear polarization} %Title of paper

% Explanatory text shou


\author{P. W{\l}odarczyk}
\affiliation{Institute of Physics, Jagiellonian University, {\L}ojasiewicza 11, 30-348, Krak\'ow, Poland}
\altaffiliation{Current affiliation: Faculty of Computer Science, Electronics and Telecommunications, AGH University of Science and Technology, Mickiewicza 30, 30-059 Krak\'ow, Poland}

\author{S. Pustelny}
\email[Corresponding author email: ]{pustelny@uj.edu.pl}
\affiliation{Institute of Physics, Jagiellonian University, {\L}ojasiewicza 11, 30-348, Krak\'ow, Poland}

\author{D. Budker}
\affiliation{Helmholtz Institute Mainz, Johannes Gutenberg University, 55099 Mainz, Germany}
\affiliation{Department of Physics, University of California at Berkeley, Berkeley, California 94720-7300, USA}
\affiliation{Nuclear Science Division, Lawrence Berkeley National Laboratory, Berkeley, California 94720, USA}

%\thanks{}

\date{\today}

\begin{abstract}
We present a technique for generating light in an arbitrary polarization state. The technique is based on interference of two orthogonally polarized light beams, whose amplitudes and phases are controlled with a Mach-Zehnder inteferometer with acousto-optic modulators (AOMs) placed in each arm. We demonstrate that via control over amplitudes, phases, and frequencies of acoustic waves driving the AOMs, any polarization state can be synthesized. In particular, we demonstrate generation of linearly polarized light, whose polarization plane continuously rotates at a rate from 1~kHz to 1~MHz. Such light finds applications in science (e.g., investigations of Bloch-Siegert effect) and technology (optically-pumped magnetometers).
\end{abstract}


\maketitle %\maketitle must follow title, authors, abstract and \pacs

\section{Introduction\label{sec:Introduction}}

Next to frequency, polarization is the most important parameter determining interaction of light with matter. Structure of molecules, orientation of molecular bonds, strength of intra- and intramolecular interactions can be determined through investigations of polarization-dependent absorption. Absorption selectivity also provides means for matter manipulation, giving rise to media with unique properties (coherently prepared media) and such phenomena as electromagnetically induced transparency,\cite{Fleischhauer2005Electromagnetically} coherent population trapping,\cite{Arimondo1996Coherent} slow light\cite{Bigelow2003Superluminal} or light storage.\cite{Phillips2001Storage} Polarization of light is also important for imaging, as specific structures of objects induce strong optical anisotropy. For example, collagen fibers have elongated structures, which make them dichroic. Since the collagen structure may degrade due to disease processes, monitoring the anisotropy may be used for biomedical purposes.\cite{Psilodimitrakopoulos2013Quantitative} Analogically, polarization rotation in chiral molecules may be used for quantitative measurements of molecules’ concentration. A particular example of such an application is determining the glucose level in blood.\cite{Boer2002Review,Lee2013Chiral} Finally, polarization is used for information encoding, where polarization-multiplaxing techniques enable doubling the capacity of telecommunication networks,\cite{Evangelides1992Polarization} while in many quantum-information protocols, polarization plays a crucial role for encoding/transferring of quantum information.\cite{Duan200Quantum,Kwiat1995New,Julsgaard2004Experimental}

Polarization of light can be easily modified with materials with naturally or externally-induced optical anisotropy. The first group includes calcite, zircon, quartz etc., while the latter encompasses electro-optical materials,\cite{Allen1996Electro} elasto-optical materials,\cite{Photoelastic1981Photoelastic} and liquid crystals.\cite{Vicari2003Optical} The situation gets more complicated when a full\footnote{Possibility of transformation of the light from any polarization state into an arbitrary polarization state.} dynamic control over the polarization is required. Since mirrors, lenses/objectives, and other optical elements are birefringent/dichroic, which also depends on environmental parameters (temperature, humidity etc.), in polarization microscopy, where well-determined polarization in a focal point is required,\cite{Lien2013Precise} active control over light polarization is a necessity. In the simplest scenario, such control can be achieved with mechanical rotation of half- and quarter-wave plates, but this solution is slow and may introduce noise, due to rotation-related mechanical vibrations, which is detrimental for some applications. Larger bandwidth may be achieved with liquid crystals, where control over the voltage applied to the crystals enables modification of polarization state of transmitted/reflected light.\cite{Rumbaugh1990Polarization,Zhuang1999Polarization} Much higher bandwidth may be obtained with a set of two electro-optical modulators where axis of one modulator is rotated by 45$^\circ$ with respect to each other.\cite{Kaneshiro2016Full} Control over the voltage applied to the modulators allows for generation of light in an arbitrary polarization state, assuming initial linear polarization of light. While this solution allows for fast modification of light polarization state, a certain disadvantage of the system is the need to reset it, which is imposed by a limited range of available voltages (high voltage needs to be dropped to the level corresponding to the same phase accumulated in the modulators, i.e., phase changes by $2\pi$ or multiples, which introduces discontinuity in polarization evolution). This prevents generation of a monotonic change of a light-polarization state, e.g., continuous rotation of linear polarization.

A new method for generating light in a given polarization state has been recently proposed.\cite{Robens2018Fast} The idea is based on the interference of two circularly polarized light beams of opposite helicities and controllable intensities and phases. This is achieved by application of acousto-optical modulators (AOMs), operating in the first order of diffraction, controlled with two high-resolution direct digital synthesis (DDS) generators. The authors demonstrated the ability to synthesize the state of polarization with high fidelity (99.99\%) and discussed the application of the technique in studies of ultra-cold gases.

In this article, we present a technique based on a complementary approach. As in Ref.~\cite{Robens2018Fast}, polarization is synthesized by overlapping two orthogonally circularly polarized beams. However, in the method of Ref.~\cite{Robens2018Fast}, a simpler polarization control system is implemented. Instead of using two 32-bit DDS generators, the AOMs are controlled with a system incorporating a single 80-MHz generator and a single-sideband modulator (SSB). The latter mixes the 80-MHz high-frequency signal (carrier) with a low-frequency (modulation) signal, generating only on the sum of the frequencies (the second sideband at the frequency difference as well as the carrier are significantly damped). The system produces two coherent light beams, whose superposition enables generation of light in an arbitrary polarization state, through control over the amplitude and phases of the AOM-driving signals. Moreover, control over the frequency difference between the two light beams, achieved with the low-frequency signal, enables continuous modification of the light polarization. This facilitates experiments, where precise and dynamic control over the light polarization is of crucial importance. For example, we demonstrate linearly polarized light with continuously rotating polarization orientation, which may find application in optical magnetometry.\cite{Budker2013Optical} A system for stabilization of parameters of generated light, simpler than that described in Ref.~\cite{Robens2018Fast}, is also demonstrated and its performance, showing significant reduction of drifts the parameters, is analyzed. 

The article is organized as follows. The next section briefly discusses theoretical grounds of the technique. Then, an experimental setup is presented and measurement results are described. Section~\ref{sec:Continuous} demonstrates generation of light with continuously changing polarization. It shows that stable evolution of the polarization can be generated with the system. Conclusions are summarized at the end of the manuscript.

\section{Principle of operation\label{sec:Principle}}

Light in an arbitrary polarization state can be generated by combining two coherent, orthogonally polarized light beams of generally different amplitudes and phases but the same frequency. While this can be achieved by a superposition of two linearly polarized light beams, application of circularly polarized light is superior both from a physical perspective (circular polarizations are eigenstates of many quantum systems) and technically (the direction of generated linear polarization is controlled through a phase shift not amplitudes of the beams). A sum of two circularly polarized light beams
\begin{equation}
\begin{aligned}
   	\vect{E}_A=&\vect{x}\frac{E_{A0}}{\sqrt{2}}\cos\omega t+\vect{y}\frac{E_{A0}}{\sqrt{2}}\sin\omega t,\\
    \vect{E}_B=&\vect{x}\frac{E_{B0}}{\sqrt{2}}\cos\left(\omega t+\varphi\right)-\vect{y}\frac{E_{B0}}{\sqrt{2}}\sin\left(\omega t+\varphi\right),
\label{eq:PolarizationComponents}
\end{aligned}
\end{equation}
where $\omega$ is the light frequency, $\varphi$ is the phase shift, $E_{A0}$ and $E_{B0}$ are amplitudes of light beams, and $\vect{x}$ and $\vect{y}$ are unit vectors, may give rise to any polarization. For example, in a specific case of a superposition of two equally intense light beams ($E_{A0}=E_{B0}=E_0$), one can generate linearly polarized light with polarization direction determined by the phase shift $\varphi$
\begin{equation}
    \vect{E}=\vect{E}_A+\vect{E}_B=E_0\sqrt{2}\left(\vect{x}\cos\frac{\varphi}{2}-\vect{y}\sin\frac{\varphi}{2}\right)\cos\omega t,
    \label{eq:ResultantPolarizatinon}
\end{equation}
where we have omitted the phase term in $\cos\omega t$. Simultaneously, control over the beam intensities provides means of tailoring the ellipticity of the light. Specifically, when intensity of one of the beams is zero ($E_{A0}=0$ or $E_{B0}$), the light is circularly polarized. Thereby, the approach enables precise control over the polarization state of light, as well as providing means of dynamic modification of the polarization state.

\section{Generation of light in an arbitrary polarization state}

A system, enabling generation of light of an arbitrary polarization state, can be implemented with a Mach-Zehnder interferometer, equipped with a part enabling control over intensities and phases of the beams propagating in each of the interferometer arms.
\begin{figure}
    \includegraphics[width=\columnwidth]{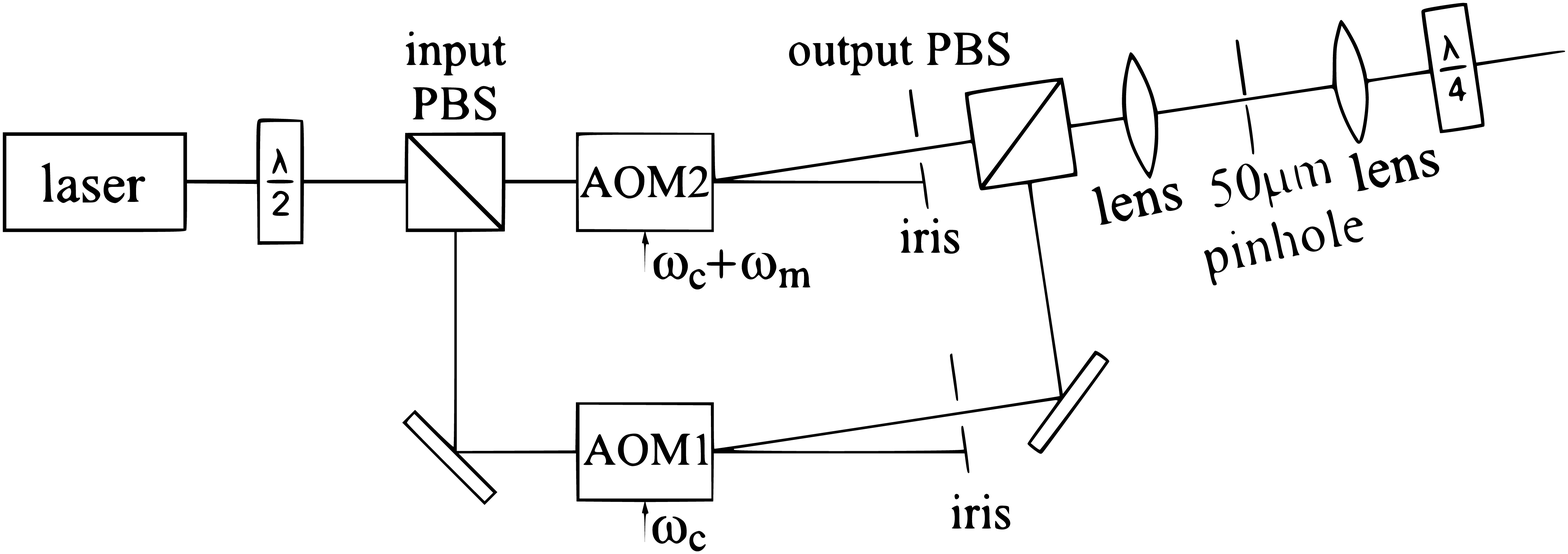}    
    \caption{Schematic diagram of the system used for generation of arbitrarily polarization light. Combination of two orthogonally polarized light beams of controllable amplitudes $E_{A0}$ and $E_{B0}$ and phase shift $\varphi(t)$ enables generation of an arbitrarily polarization state. This is implemented by application of a half-wave plate, two acousto-optical modulators (AOM1 and AOM2), driven by phase coherent electric signals of frequencies $\omega_{AOM1}$ and $\omega_{AOM2}$, and a quarter-wave plate ($\lambda/4$) situated at the end of the system. Additionally, the phase of the $\omega_{AOM1}$ signal can be varied. PBS abbreviates a polarizing beam splitter.}
    \label{fig:Setup}
\end{figure}
In the system, linearly polarized laser light passes through a half-wave plate, which provides roughly equal splitting of light intensity at the input polarizing beam splitter (PBS). Orthogonally polarized light beams, directed into the interferometer arms, pass through AOMs, operating in the first order of diffraction. The diffracted light is frequency shifted by the frequency of acoustic waves. The diffracted beams are then combined at the output PBS and transmitted through a lens-pinhole system,\footnote{This can be replaced with a single-mode polarization-maintaining optical fiber, where polarization of the beams are oriented along the fiber axes.} which ensures spatial overlap of the two beams. The last element in the system is a quarter-wave plate ($\lambda/4$), rotated by 45$^\circ$ with respect to the polarization of both beams, which transforms linear polarizations of the beams into their circular counterparts (left and right circularly polarized light). The superposition of the two beams results in generation of light in a polarization state controlled through amplitudes and phases of the interfering beams. The control over these parameters is obtained by modifying the amplitudes and phases of the electric signals driving the AOMs. In our experiments, the signals come from the same (80~MHz) generator but their amplitudes are independently controlled. Additionally, the phase of one of the signals (carrier) can be varied. Changing the amplitude of the electric signal controls efficiency of the beam diffraction in the AOM (control of light intensity) and hence provides control over ellipticity of the output light. The control over the phase difference between the two signals allows for modification of the spatial orientation of the polarization. Crucial in this context is the AOM's ability to imprint an electric/acoustic wave parameters onto the parameters of diffracted light. In such a case, modification of signal phase changes the phase difference $\varphi$ between the two interfering beams.

An important issue stemming from the application of a Mach-Zehnder interferometer is that the phase difference between the interfering beams not only depends on the phase shift between AOM driving signals, but also on the optical lengths of interferometer arms. Therefore, to obtain a particular polarization state an additional offset needs to be added to the AOM driving signal to compensate for the imbalance of the optical lengths of the arms. Importantly, the difference is not constant over time but drifts (for example, due to thermal changes of the interferometer size, mechanical vibration of the interferometer, and air movement) resulting in uncontrolled modification of the polarization state. Therefore, to achieve reliable operation of the system, an additional module monitoring the polarization state and correcting it for its changes needs to be implemented. This can be done, for example, by monitoring the Stokes parameters of the wave and imprinting additional phase on one of the driving signals. The parameters can be detected with a set of polarizing and nonpolarizing beam splitters, wave plates, and balanced photo-detectors,\cite{Goldstein2011Polarized} which, through a feedback loop, are used to correct for the changes in the interferometer geometry. Such a system was used in our setup providing stability of generated polarization.

\begin{figure}
	\includegraphics[width=\columnwidth]{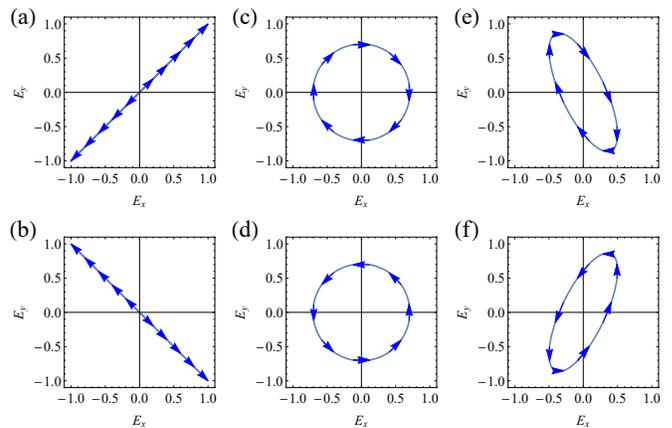}
	\caption{Polarization states. (a) $E_{A0}/E_{B0}=1$, $\varphi=0$, (b) $E_{A0}/E_{B0}=1$, $\varphi=\pi$, (c) $E_{A0}/E_{B0}=\infty$, $\varphi=\pi/2$, (d) $E_{A0}/E_{B0}=0$, $\varphi=\pi/2$, (e) $E_{A0}/E_{B0}=(1+\sqrt{3})/(1-\sqrt{3})$, $\varphi=3\pi/4$, and (f) $E_{A0}/E_{B0}=(1+\sqrt{3})/(1-\sqrt{3})$, $\varphi=-\pi/4$.\label{fig:PolarizationState}}
\end{figure}

Examples of polarization of synthesized light generated with the system and measured with a polarization analyzer are shown in Fig.~\ref{fig:PolarizationState}.
The first two plots [Fig.~\ref{fig:PolarizationState}(a) and (b)] depict linearly polarized light. Such polarization arises when same intensities but different phases of two interfering beams are used. In particular, the two shown polarization states were obtained when the phase difference was set to 0$^\circ$ and $180^\circ$. The circular polarizations [Fig.~\ref{fig:PolarizationState}(c) and (d)] are obtained with only one of the beams (no driving signal was applied to one of the AOMs). In such a case, the linearly polarized light, propagating in a given channel of the interferometer, is transformed into circularly polarized light of a specific helicity. Finally, the two examples of elliptically polarized light beams are shown in Fig.~\ref{fig:PolarizationState}(e) and (f). In this case, the amplitudes, as well as, phases of the two signals are different [by $180^\circ$ between (e) and (f)]. In accordance with former cases, the amplitude ratio of two beams determines the ellipticity of the beam, while phase difference controls the alignment of the principal axes of the polarization ellipse.

\section{Generation of light with continuously rotating linear polarization\label{sec:Continuous}}

While generation of light in an arbitrary polarization state can be implemented in various schemes,\cite{Robens2018Fast,Kaneshiro2016Full,Moreno2011Complete,Clegg2013Double} a unique feature of the system described here is its ability to generate light with continuously/monotonically varied polarization state. A specific example is linearly polarized light with continuously precessing polarization. Generation of such light opens new means for investigations of light-atom interactions and its application, e.g., magnetometry.

Equation~\eqref{eq:ResultantPolarizatinon} shows that the alignment of the linear-polarization axis depends on the phase difference between the beams. Therefore, to generate light with rotating linear polarization, one needs to continuously increase the phase difference between the two beams. This can be done by offsetting the frequency difference of the two beams. In such a case, the phase between the beams is steadily increased, which results in polarization rotation. Based on Eq.~\eqref{eq:PolarizationComponents}, it can be shown that superposition of two orthogonally polarized light beams of opposite helicity, shifted in frequency by $\Delta\omega$, corresponds to
\begin{equation}
	\vect{E}=E_0\sqrt{2}\left[\vect{x}\cos\left(\frac{\Delta\omega}{2}t\right)-\vect{y}\sin\left(\frac{\Delta\omega}{2}t\right)\right]\cos\omega t,
	\label{eq:RotationPolarization}
\end{equation}
where $\omega\gg\Delta\omega$. As a result, linear polarization of light rotates at half of the difference frequency $\Delta\omega/2$.

In principle, generation of two electric signals shifted in frequency can be implemented with two independent, clock-referenced generators, such as in Ref.~\cite{Robens2018Fast}. This solution, however, has several drawbacks. The first is limited control over the frequency of the driving signal. Particularly, generation of 80~MHz (a driving frequency of many AOMs) with hertz or subhertz resolution is possible but such generators are expensive. In turn, control over the polarization rotation frequency at the subhertz level is difficult and expensive. Moreover, in many schemes, control of low-frequency signal, corresponding to the difference of the two high-frequency signals, is superior. This is the case because stabilization of low-frequency signal and hence the frequency difference at subhertz level is much simpler. Finally, even with clock-referenced generators their frequency difference typically drifts at the level of 10$^{-7}$~s$^{-1}$ or larger, which introduces additional instabilities into the polarization of generated light.

\begin{figure}
	\includegraphics[width=1\columnwidth]{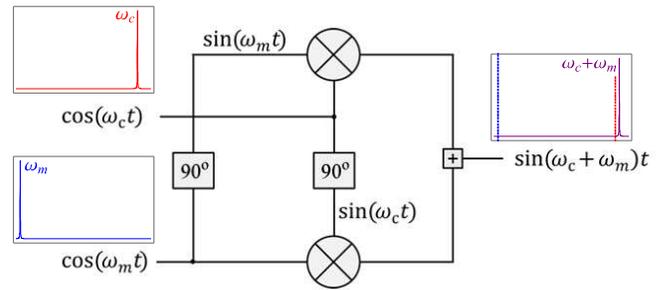}
	\caption{Single-sideband modulator used to generate signals at a sum of two signals (the carrier and the modulation signal). The signal is synthesized by phase shifting two highly spectrally different input signals (many orders of magnitude difference), mixing them with original signals and finally summing the results of the mixing.\label{fig:SSB}}
\end{figure}

To address these issues, we used a single-sideband (SSB) modulator in a Hartley configuration (see, for example, Ref.~\cite{Nelson2007High}]. This system combines two significantly spectrally different signals (the carrier and the modulation signal), generating a signal at the sum of their frequencies. In this arrangement, the carrier and signal at the difference frequency are strongly suppressed. From a mathematical standpoint, the modulator realizes an operation given by the left-side of the equation 
\begin{equation}
	\sin\omega_ct\cos\omega_mt+\cos\omega_ct\sin\omega_mt=\sin[(\omega_c+\omega_m)t],
	\label{eq:Trigonometry}
\end{equation}
where $\omega_c$ is the carrier frequency and $\omega_m$ is the modulation frequency. As Eq.~\eqref{eq:Trigonometry} is a trigonometrical identity, the result of the operation is a signal at the sum of $\omega_c$ and $\omega_m$. Electronically, this is done by superimposing appropriately phased high-frequency carrier with low-frequency modulation (Fig.~\ref{fig:SSB}).
In the system, both carrier and modulation signals are split and shifted by 90$^\circ$. Multiplication of the signal and its successive summation, in accordance with Eq.~\eqref{eq:Trigonometry}, enables generation of the summed frequency signal. Suppression of carrier and unwanted sideband depends on the quality of the frequency shifters and linearity of the multiplexers and summer. In our experimental apparatus the undesired frequency components were suppressed by 2-3 orders of magnitude. 

Application of the SSB modulator is superior to conventional techniques of mixing signals, because it does not lead to generation of the output signal with a spectrum consisting of at least carrier and more sidebands (if modulation is nonsinudoidal many more sidebands are present in the output).

\begin{figure}
	\includegraphics[width=\columnwidth]{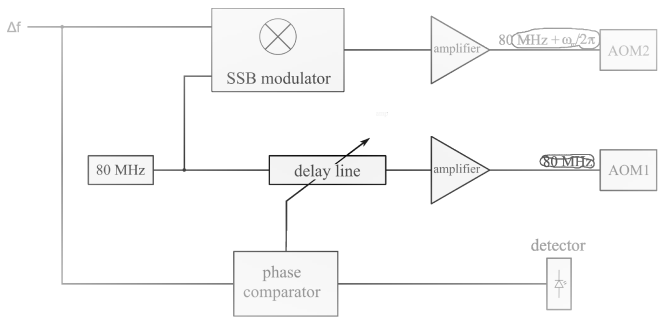}
	\caption{Electronic circuit driving the AOMs. In the upper channel 80-MHz signal is frequency shifted with an SSB modulator to generate rotating polarization. In the lower one it is phase shifted to adjust the phase of this rotation. The detector monitoring the polarization and phase comparator form a feedback loop which stabilizes jitter of the rotation caused by air movements in the interferometer's arms.\label{fig:Electronic}}
\end{figure}

The circuits used in the SSB modulator as 90$^\circ$ phase shifters can only operate properly in a finite bandwidth. Typically, the range of frequencies $\Delta\omega_m$ supporting reliable operation of the SSB modulator lies within one decade. Thereby, the offset frequency, determined by $\omega_m$ of the low-frequency signal, and hence the polarization-rotation frequency, cannot be arbitrarily small. This may be a problem due to slow drifts of optical lengths of the interferometer arms and thus the phase of the polarization rotation. To address this issue, we implemented a variable delay line in electronics driving the 80-MHz AOM (Fig.~\ref{fig:Electronic}). The line allows us to change the phase of the driving electric signal in a wide range (over 4$\pi$) and thus compensate for the phase drift using a feedback loop. In our scheme, an error signal is supplied by the phase comparator, which stabilizes the phase of the low-frequency signal with the phase of the rotation obtained from an additional detector placed in the resultant beam path. The detector consists of a linear polarizer placed in front of a photodiode. In such an arrangement, rotating linear polarization produces sinusoidally modulated photocurrent at twice the modulation frequency $\omega_m$,\footnote{Note that the rotation occurs at $\omega_m/2$ [Eq.~\eqref{eq:RotationPolarization}].} which is fed into the comparator. The implemented integrator system controls the phase of the 80-MHz signal keeping the low-frequency signal and the reference signal phased. If phase shift is too large ($>\!\!4\pi$), then the system automatically zeros the phase difference by zeroing the voltage in the delay line. This phase reset is the only discontinuity in operation of our system, and may occasionally cause some instantaneous unwanted polarization states. It is noteworthy that the phase of the reference signal can be controlled by rotating the polarizer, which allows us to zero the phase difference but also to introduce a static phase shift between the driving signal and the rotation.

\begin{figure}
	\includegraphics[width=0.9\columnwidth]{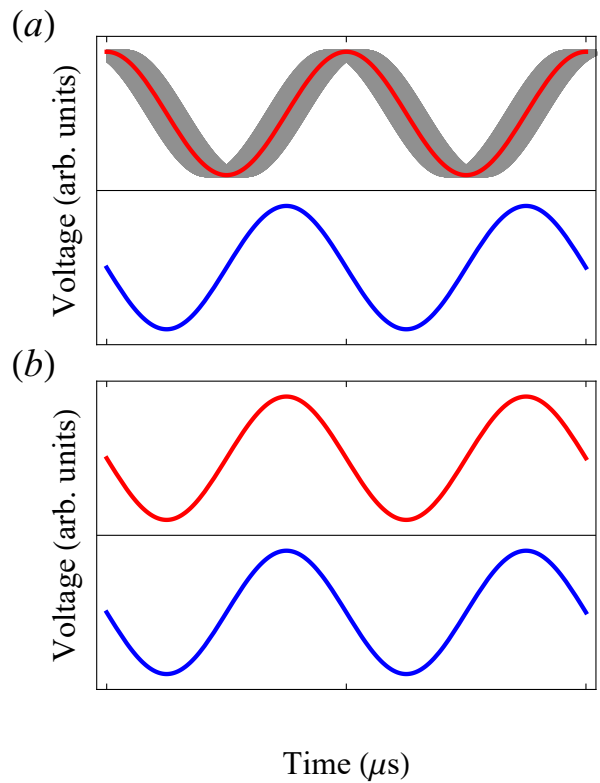}
	\caption{Rotation signal (top red traces) and reference signal (lower blue traces) observed without (a) and with (b) phase stabilization. For the unlocked system, the phase of the rotation signals drifts by about $\pm 30^\circ$ within 10~s, despite isolating the interferometer from its surrounding by enclosing it in a cardboard box. When the stabilization system was turned on, the drift was significantly reduced. The signals were measured with 1-MHz modulation signal.\label{fig:Stability}}
\end{figure}

Figure~\ref{fig:Stability} compares the polarization-rotation signal [red traces in (a) and (b)] with the low-frequency signal used for frequency shift (blue traces in lower panels) operating without (a) and with (b) the phase stabilization.
Without the stabilization, the phase difference between the rotation and reference signal is random and drifts over time. The shaded area in (a) shows the drifts observed within about 10 seconds (about $\pm 30^\circ$). Such a large drift is observed despite the fact that the interferometer was enclosed inside a cardboard box, isolating the interferometer from unwanted air motions. No such drift is observed when the stabilization system is turned on. Moreover, in this scheme the phase between the two signals is locked.

To assess the stabilization more quantitatively, we measured a phase of the rotation signal with respect to the reference signal using a lock-in amplifier (Stanford Research SR844). The signal was measured with a 10-ms time constant and 24-dB/oct filter. 
\begin{figure}
	\includegraphics[width=0.9\columnwidth]{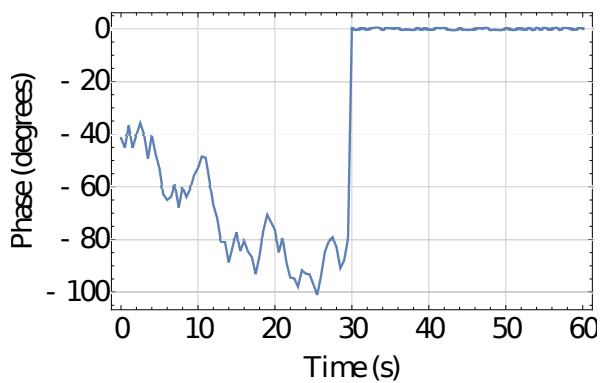}
	\caption{Phase difference between driving signal and rotation signal measured with ($t<30$~s) and without ($t>30$~s) active stabilization of the phase. After turning on the stabilization system, the phase locks at 0$^\circ$ and its fluctuations are reduced by nearly two orders of magnitude.\label{fig:Drifts}}
\end{figure}
The results presented in Fig.~\ref{fig:Drifts} show that turning on the stabilization system results in stabilization of the phase at a given value (0$^\circ$ in the case) and significant reduction of the noise. The fluctuations and drift of the signal are reduced by nearly two orders of magnitude (root mean square of the phase signal reduces from 15$^\circ$ observed in the first half of the measurement to below 0.3$^\circ$ in the second half). This example demonstrates the significance of the stabilization system, which is crucial for many applications.

\section{Summary}

We have described a system enabling synthesis of light in an arbitrary polarization state. By interfering two light beams of orthogonal circular polarizations, we demonstrated generation of light of arbitrarily oriented linear polarization, circular polarizations of opposite helicities and elliptically polarized light.

In addition to generation of light of an arbitrary polarization state, we also demonstrated the ability of the system to generate light with continuously changing polarization. This was shown by synthesizing linearly polarized light, whose linear polarization rotated around the propagation direction. Control over the precession frequency, implemented using a single-sideband modulator, within a broad frequency range (from 1~kHz to 1~MHz), opens avenues for application of the technique for optical pumping. In this context, the technique may be used for efficient generation of optical anisotropy in media subjected to an external magnetic field. Such an approach is used in optical magnetometers \cite{Budker2013Optical} exploiting nonlinear magneto-optical rotation.\cite{Budker2002Resonant} Application of rotating polarization to optical magnetometry will be presented elsewhere.

\begin{acknowledgments}

The authors would like to express their gratitude to David S. Weiss, Derek Jackson Kimball, Valeriy Yashchuk, Alexander Akulshin, and Antoine Weis for stimulating discussions and Irena Rodzo\'n and Julia Sudyka for help in construction and optimization of experimental apparatus. This work was supported by the grant number 2015/19/B/ST2/02129 financed by the Polish National Science Centre.

\end{acknowledgments}

\bibliography{SystemFor}

\end{document}